\documentclass{article}
\usepackage{spconf,amsmath,graphicx,hyperref}
\usepackage{multirow}  
\usepackage{booktabs}
\usepackage{makecell} 

\title{Specific Multi-emitter Identification via Multi-label learning}
\name{Yuhao Chen, Boxiang He, Shilian Wang, Jing Lei}
\address{College of Electronic Science and Technology, National University of Defense Technology, China}
\newtheorem{remark}{Remark}
\begin{document}
%
\maketitle
\begin{abstract}
Specific  emitter identification leverages hardware-induced impairments to uniquely determine a specific transmitter. However, existing approaches fail to address scenarios where signals from multiple emitters overlap. In this paper, we propose a specific multi-emitter identification (SMEI) method via multi-label learning to determine multiple transmitters. Specifically, the multi-emitter fingerprint extractor is designed to  mitigate the mutual interference among overlapping signals. Then, the  multi-emitter decision maker is proposed to assign the all emitter identification using the previous extracted  fingerprint. Experimental results demonstrate that, compared with baseline approach, the proposed SMEI scheme achieves comparable identification accuracy under various overlapping conditions, while operating at significantly lower complexity. The significance of this paper is to identify multiple emitters from overlapped signal with a low complexity.

\end{abstract}
\section{Introduction}
\label{sec:intro}

Specific  emitter identification (SEI) uses inherent hardware imperfections to identify devices at the physical layer \cite{Sankhe2019ORACLE}. Most SEI studies emphasize individual identification and overlook concurrent device transmissions \cite{Zheng2023TargetIdentification,Wei2024SourceIdentification}, despite multiple emitters operating simultaneously. For example,  the industrial, scientific, and medical (ISM) band is often subject to simultaneous emissions from multiple devices. These emitters typically occupy adjacent sub-bands and transmit within closely spaced time intervals, leading to energy superposition that manifests as spectral overlap and multi-source signal mixtures \cite{Zeng2019UAV5G}. In practical scenarios, the received signal comprises a mixture of emissions from multiple sources, motivating the extension of SEI to specific multi-emitter identification (SMEI), which aims to identify the entire set of active devices rather than a single emitter.

Enumerating all possible active combinations leads to exponential growth of the theoretical output space as the number of devices 
$K$ increases, since the number of non-empty subsets equals $2^{K}-1$. This exponential inflation not only escalates training and inference costs, but also introduces data-combination sparsity and degrades generalization to previously unseen combinations \cite{Chen2021RFMixtureDDL}. Furthermore, receiver noise can severely contaminate the subtle emitter-specific distortion features, making SMEI particularly challenging \cite{Jin2023RFFeatureFusion}.

In multi-label classification, each input sample can simultaneously belong to multiple classes, unlike traditional single-label classification where each sample is associated with only one class. Multi-label classification can significantly reduce the number of model parameters by reducing the number of combinations to consider, thus decreasing the consumption of computational resources during training and inference \cite{Li2023CompoundJamming}. This framework naturally models scenarios such as multi-emitter identification, where a received signal may contain contributions from several active emitters, requiring the system to assign multiple labels to a single observation. Here, we propose a SMEI method via multi-label learning. Specifically, we first model the SMEI problem. Then, we design a multi-emitter feature extractor to capture multi-emitter features, followed by a decision-making process that assigns emitter labels based on the extracted features. Finally, numerical results show that our scheme has the promising performance in terms of the identification accuracy and the complexity. The main contribution of this paper is the introduction of multi-label learning to address the challenging SMEI problem with extremely low complexity.

\section{System Model}
\label{sec2}

\begin{figure*}[t]
	\centering
	\includegraphics[width=0.7\textwidth]{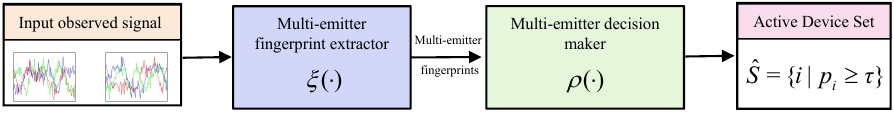}
	\caption{SMEI scheme via multi-label learning. }
	\label{fig:smei_framework}
\end{figure*}
\begin{figure*}[htbp]
	\centering
	\includegraphics[width=0.7\textwidth]{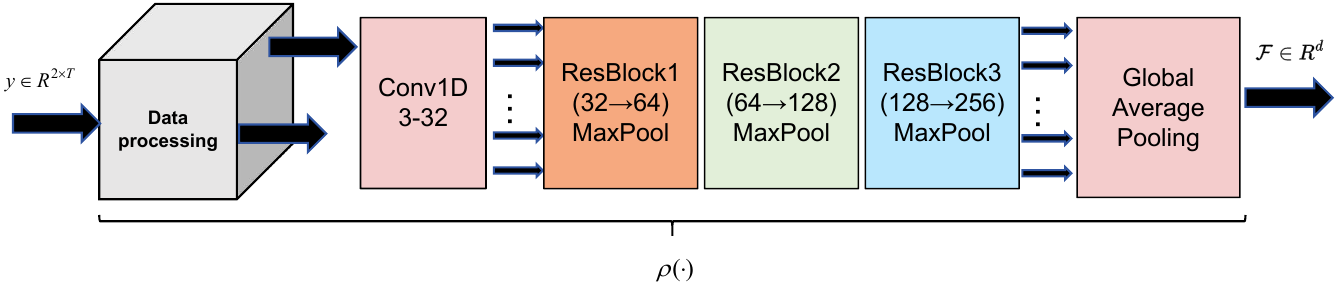}
	\caption{Schematic diagram of the multi-emitter fingerprint extractor. }
	\label{fig:MEFE_framework}
\end{figure*}

We consider up to \(K\) potential emitters, where the received signal is  denoted by
\begin{equation}
	y \;=\; \sum_{m\in S} h_m \, f_m\!\bigl(x_m\bigr) \;+\; w,
	\label{eq:rx_overview}
\end{equation}
where \(S\) is the  set of active devices; \(x_m\) denotes the original transmitted of the $m$-th emitter; \(f_m(\cdot)\) is the distortion function of the $m$-th emitter; \(h_m\) denotes the channel coefficient; \(w\) is Gaussian noise. 

Following the literature \cite{He2020CooperativeSEI}, we model the distortion function \(f_m(\cdot)\) using the I/Q imbalance, the spurious tone, the carrier leakage, and the power amplifier (PA) nonlinearity. Specifically,  the distorted signal with the I/Q imbalance is expressed as
\begin{equation}
	x_m'=\; \mu_m\, x_m \;+\; \nu_m\, x_m^{*},
	\label{eq:iq_model_new}
\end{equation}
where $\mu_m$ and $\nu_m$ are the parameters used to describe the distortion of the modulator, which can be represented as
%
%
%
%

\begin{equation}
	\mu_m = \tfrac{1}{2}(G_m+1)\cos\left(\tfrac{\zeta_m}{2}\right) + j\,\tfrac{1}{2}(G_m-1)\sin\left(\tfrac{\zeta_m}{2}\right),
	\label{eq:mu_m}
\end{equation}

\begin{equation}
	\nu_m = \tfrac{1}{2}(G_m-1)\cos\left(\tfrac{\zeta_m}{2}\right) + j\,\tfrac{1}{2}(G_m+1)\sin\left(\tfrac{\zeta_m}{2}\right),
	\label{eq:nu_m}
\end{equation}
where $\zeta_m$ is the phase bias and $G_m$ denotes the gain imbalance. With the spurious tone and carrier leakage,  the distorted signal is further formulated by 
\begin{equation}
	x_m'' \;=\; x_m' e^{j 2\pi f t}\;+\; a_m^{ST}\, e^{j{2\pi (f_m+f_m^{ST})}\,}\;+\; \zeta_m\, e^{j2\pi f t\,},
	\label{eq:if_spur_cl_new}
\end{equation}
where \( a_m^{ST} \) and \( f_m^{ST} \) are the amplitude and the frequency of the spurious tone; \( \zeta_m e^{j 2 \pi f t} \) is the carrier leakage.

Finally, the signal $x_m''$ is fed into the PA with the nonlinear distortion, which can be expressed as

\begin{equation}
	x_m''' = \sum_{l=1}^{L} b_{l,m}\, \left( x_m'' \right)^{\,l},
	\label{eq:pa_taylor_first}
\end{equation}

\begin{equation}
	= f_m(x_m),
	\label{eq:pa_taylor_second}
\end{equation}
where \( L \) is the order and \( b_{m,l} \) denotes the coefficient of the taylor polynomial. The goal of this paper is to determine the multiple emitters from the received overlapped signal $y$.

\section{Proposed SMEI Scheme via Multi-label learning}
\label{sec:format}
The core idea of our SMEI scheme is to simultaneously identify multiple active emitters using a multi-label learning. Figure~\ref{fig:smei_framework} presents the block diagram of the proposed SMEI scheme. The process starts with the overlapped signal \( y \), followed by the multi-emitter fingerprint extractor \( \xi(\cdot) \), which obtains multi-emitter fingerprints. The features are then input to the multi-emitter decision maker $\rho(\cdot)$, resulting in a set of active emitters \( S \).

\subsection{Multi-emitter fingerprint extractor}

We design a multi-emitter fingerprint extractor \( \xi(\cdot) \), which aims to extract multi-emitter fingerprints from received overlapped signals for subsequent multi-label classification tasks. The obtained multi-emitter fingerprints can be represented as
\begin{equation}
\mathcal{F} = \xi(y) : R^{2 \times T} \to R^{d},
\end{equation}
where \( y \in R^{2 \times T} \) indicates the overlapped complex signal with a length of \( T \) samples; \( d \) defines the dimensionality of the extracted feature space.
Figure \ref{fig:MEFE_framework} illustrates the architecture of the fingerprint feature extractor, which employs convolutional layers, residual blocks, and pooling layers to achieve multi-scale feature extraction.  

Specifically, the overlapped signal \( y \) is converted into its real and imaginary parts after data preprocessing, resulting in an input of shape \(\mathcal{R}^{2 \times T}\). The input is then expanded to 32 dimensions through convolutional layers and progressively increased to 256 dimensions via multiple residual blocks. This process enhances feature representation through skip connections and mitigates the vanishing gradient problem. The max pooling layers incrementally reduce the size of the feature maps. Global average pooling compresses the feature space from \((B, 256, L/8)\) to a fixed-dimensional feature vector \((B, 256)\), where \(B\) is the batch size. Finally, \(\mathcal{F}\) effectively captures the multi-emitter fingerprint information from \( y \).

\subsection{Multi-emitter decision maker}

Multi-emitter decision maker $\rho(\cdot)$ is to map the feature space \( R^{d} \) to emitter state space, thereby effectively identifying up to \( K \) active emitters. Formally, we define the label vector $\boldsymbol{\lambda} = \begin{bmatrix} \lambda_1 & \lambda_2 & \cdots & \lambda_K \end{bmatrix}^{\top}$, where $\lambda_m \in \{0,1\}$ indicates the active state of the $m$-th emitter.

The mapping relationship of the multi-emitter decision maker $\rho(\cdot)$ can be expressed as \( \rho: R^{d} \to 2^{[1,K]} \), where the input is multi-emitter fingerprints \( \mathcal{F} \in R^{d} \) and the output is the estimated active emitter set \( \hat{S} \).  Thus, the set of active emitters recognized by the SMEI can be expressed as 

\begin{equation}
	\hat{S} = \{m \mid \sigma(\rho(\mathcal{F}))_m > \theta, m \in [1,K]\},
\end{equation}
where \( \sigma \) denotes the sigmoid activation function.
We use the binary cross-entropy loss function, which calculates the error between the predicted activation probabilities and the true labels for each emitter, i.e.
\begin{equation}
	\mathcal{L}_{\text{BCE}} = -\frac{1}{K} \sum_{m=1}^{K} \Big[ \lambda_m \log(p_m) + (1 - \lambda_m) \log(1 - p_m) \Big],
	\label{eq:bce_loss}
\end{equation}
where

\begin{equation}
	p_m = \sigma(\rho(\mathcal{F}))_m,
\end{equation}
where $p_m$ is the predicted activation probability for the $m$-th emitter; $\sigma(\cdot)$ is the sigmoid activation function.

\begin{remark}
The proposed SMEI scheme overcomes single-label limits by using multi-label learning and independent sigmoid modeling for each emitter, accurately capturing activation in signal-overlap scenarios while maintaining identification accuracy and reducing system complexity.

\end{remark}

\subsection{Evaluation metric for SMEI}

To evaluate the performance of the proposed SMIE sheme, we adopt two key accuracy metrics, subset accuracy \cite{DeSilva2025EVTC} and hamming accuracy \cite{Poonia2024HEA}, which is expressed as

\begin{equation}
	P_c^{\text{subset}} = \frac{1}{N} \sum_{i=1}^{N} \mathcal{I}\left(\boldsymbol{\hat{\lambda}}^{(i)} = \boldsymbol{\lambda}^{(i)}\right),
\end{equation}

\begin{equation}
	P_c^{\text{Hamming}} = \frac{1}{K} \sum_{k=1}^{K} \frac{1}{N} \sum_{i=1}^{N} \mathcal{I}\left(\boldsymbol{\hat{\lambda}}_k^{(i)} = \boldsymbol{\lambda}_k^{(i)}\right),
\end{equation}
where $N$ is the total number of samples; $\boldsymbol{\hat{\lambda}}^{(i)}$ is the predicted label vector for the $i$-th sample; $\boldsymbol{\lambda}^{(i)}$ denotes the ground truth label vector for the $i$-th sample; $\boldsymbol{\hat{\lambda}}_k^{(i)}$ is the predicted label for the $k$-th emitter of the $i$-th sample; $\boldsymbol{\lambda}_k^{(i)}$ denotes the ground truth label for the $k$-th emitter of the $i$-th sample; $\mathcal{I}(\cdot)$ is the indicator function that returns one if the condition is true and zero otherwise.

In addition, we introduce the macro F1 score as an auxiliary metric, which can be expressed

\begin{equation}
	F1_{\text{macro}} = \frac{1}{K} \sum_{k=1}^{K} \frac{2 \text{TP}_k}{2 \text{TP}_k + \text{FP}_k + \text{FN}_k},
\end{equation}

\noindent where $\text{TP}_k$ is the number of true positives; $\text{FP}_k$ denotes the number of false positives; $\text{FN}_k$ is the number of false negatives for the $k$-th class.

Here, we emphasize the characteristics of these three metrics. $P_c^{\text{subset}}$ focuses on the accuracy of the overall multi-emitter combination identification, while $P_c^{\text{Hamming}}$ provides a fine-grained analysis of the identification precision for individual emitters. However, since $P_c^{\text{subset}}$ is extremely stringent for the proposed SMEI scheme, directly comparing $P_c^{\text{subset}}$ of the two approaches is not fair. To achieve a more just and comprehensive performance comparison, this paper introduces $F1_{\text{macro}}$ as the auxiliary metric for evaluating the performance differences between the proposed SMEI scheme and prior multi-classification scheme.


\section{Experiments and Analysis}  
\label{sec:NUMERICAL RESULTS}

\subsection{Experimental setup}
This section outlines the key experimental configuration used for synthesizing the dataset, including signal modulation, spectrum allocation, device hardware parameters, and channel/noise models. The data synthesis targets the drone communication band from 2.40 to 2.48 GHz, utilizing quadrature phase shift keying modulation. To mitigate the boundary transients introduced by pulse shaping, the model input uniformly selects steady-state time intervals in the middle of each sequence. Table \ref{tab:combined} lists the  configuration parameters for the SMEI system, encompassing signal parameters, training hyperparameters and optimizer settings. Regarding spectrum allocation and digital intermediate frequency configurations, the experiment includes two levels of spectrum overlap: 50\% and 100\%. With 50\% overlap, neighboring carrier frequencies are separated by 13 MHz, which is half the symbol bandwidth. Furthermore, emitter distortion parameters are detailed in \cite{He2020CooperativeSEI} and unless otherwise specified the channel used by default is the rician channel.

\begin{table}[t]
	\centering
	\caption{System configuration parameters in multi-emitter identification experiments.}
	\label{tab:combined}
	\resizebox{\linewidth}{!}{%
		\begin{tabular}{llp{3cm}}
			\toprule
			\textbf{Category} & \textbf{Parameter} & \textbf{Value / Range} \\
			\midrule
			\multirow{6}{*}{Signal parameters} 
			& Sampling rate & \(F_s\) = 120 MHz \\
			& Symbol rate & \(R_s\) = 20 MHz \\
			& Oversampling ratio & \textit{SPS} = \(F_s / R_s = 6\) \\

			& RRC roll-off & \(\alpha\) = 0.3 \\
			& RRC span & \textit{span} = \(10 \cdot\) (symbols) \\
			& Rician factor & \(K_R\) = 10 dB \\
			\midrule
			\multirow{5}{*}{Training hyperparameters} 
			& Batch size & 64 \\
			& Epochs & 100 \\
			& Learning rate & $3 \times 10^{-4}$ \\
			& Step size & 20 \\
			& Learning rate decay factor & 0.5 \\
			\midrule
			\multirow{3}{*}{Optimizer}
			& Type & Adam \\
			& Initial learning rate & $3 \times 10^{-4}$ \\
			& Learning rate scheduler & StepLR \\
			\bottomrule
		\end{tabular}%
	}
\end{table}

\subsection{Experimental results}

\begin{figure}[t]
	\centering
	\includegraphics[width=7.65cm]{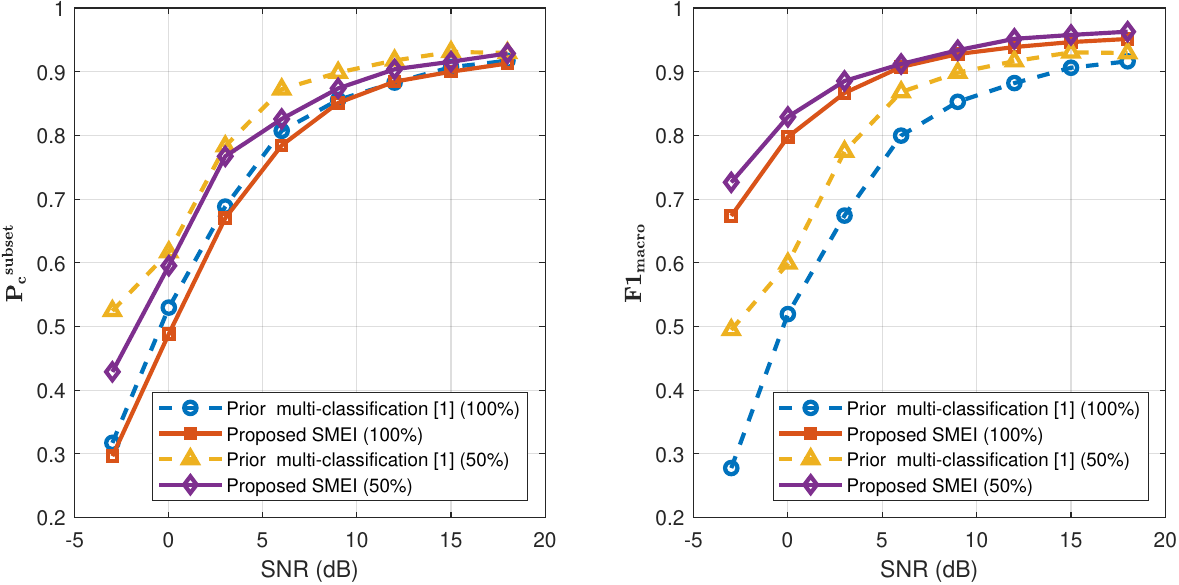}
	\caption{Comparison of $P_c^{\text{subset}}$ and $F1_{\text{macro}}$ between the proposed SMEI and prior multi-classification scheme under different overlap levels when $K = 3$.}
	\label{result1}
\end{figure}

\begin{figure}[t]
	\centering
	\includegraphics[width=4.65cm]{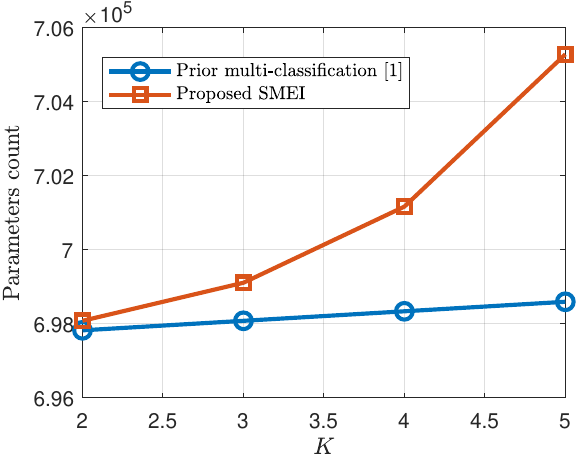}
	\caption{Comparative analysis of parameters count in the proposed SMEI and prior multi-classification scheme across varying $K$. }
	\label{result2}
\end{figure}


As shown in Figure \ref{result1}, We evaluate the performance of the proposed SMEI scheme in comparison with prior multi-classification scheme with $K = 3$. Under $100\%$ overlap conditions, as the signal-to-noise ratio (SNR) increases from $-3$ dB to $18$ dB, in the prior multi-classification scheme,  $P_c^{\text{subset}}$ increases from $0.3171$ to $0.9171$, while $F1_{\text{macro}}$ improves from $0.2775$ to $0.9163$. In our SMEI scheme, $P_c^{\text{subset}}$ rises from $0.2962$ to $0.9129$, approaching the performance of the multi-classification scheme, while $F1_{\text{macro}}$ quickly jumps from $0.6735$ to $0.9514$, showcasing outstanding macro-level identification performance. Under $50\%$ overlap conditions, the performance is further enhanced. Moreover, SMEI's identification accuracy significantly exceeds random guessing at a low SNR. This indicates strong feature extraction capabilities for effective multi-emitter identification, and as the SNR increases, identification performance improves substantially.

Figure \ref{result2} illustrates parameters count of the SMEI and prior multi-classification scheme across varying $K$ values. The SMEI scheme shows a linear increase in parameters, growing from 697,826 at $K=2$ to 698,597 at $K=5$, with an increase of just 771 parameters. This indicates high parameter efficiency and low sensitivity to emitter count. In contrast, the multi-classification scheme experiences significant growth, jumping from 698,083 to 705,279 parameters (a 7,196 increase) as $K$ rises from 2 to 5, suggestive of exponential growth. As $K$ scales up, the multi-classification scheme risks catastrophic parameter inflation, leading to exponential increases in computational and storage requirements. Therefore, we can conclude that the advantage in lower parameter count makes SMEI easier to deploy.

\begin{figure}[t]
	\centering
	\includegraphics[width=6.5cm]{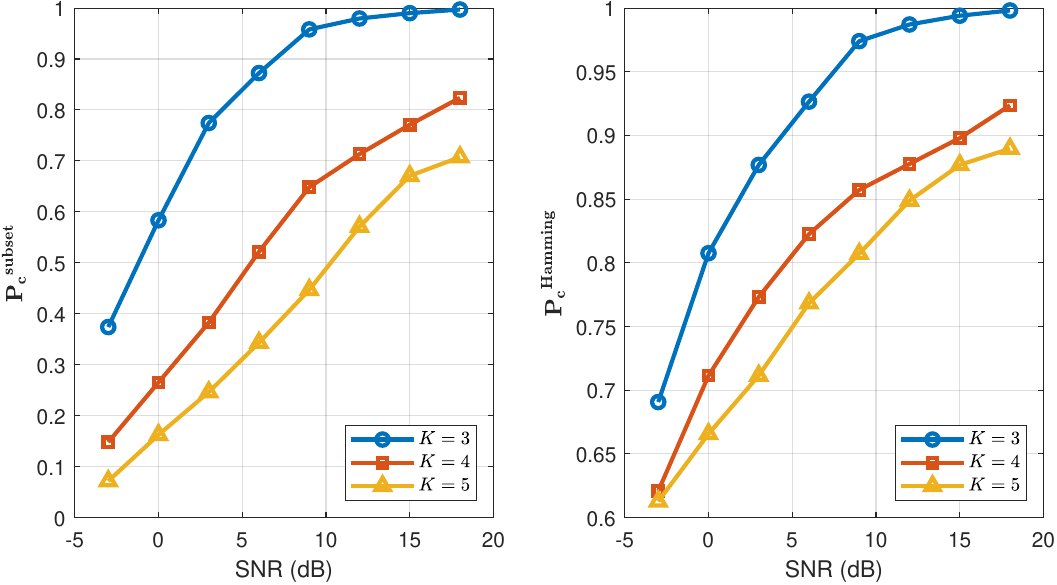}
	\caption{$P_c^{\text{subset}}$ and $P_c^{\text{Hamming}}$ of proposed SMEI for different values of $K$ in an AWGN channel with 100\% overlap. }
	\label{result3}
\end{figure}

Figure \ref{result3} presents the $P_c^{\text{subset}}$ and $P_c^{\text{Hamming}}$ of the proposed SMEI scheme for different values of $K$ in an additive white gaussian noise (AWGN) channel with 100\% overlap. Specially, when \( K=3 \), \( P_c^{\text{subset}} \) under the AWGN channel is significantly higher than the corresponding value shown in Figure \ref{result1} for the rician channel. This difference can be attributed to the inherent characteristics of the channels, where the AWGN channel typically provides a more stable and favorable environment for signal processing compared to the multipath effects present in the rician channel. Moreover as $K$ increases the impact on $P_c^{\text{subset}}$ becomes more significant. When \( K=3 \), the identification accuracy nearly reaches complete identification at an SNR of 18 dB. However, for \( K=5 \), \( P_c^{\text{subset}} \) is approximately 0.7074, indicating that as \( K \) increases, subset-level identification becomes progressively more challenging. In contrast, the bit-wise accuracy based on hamming indicates that most identification errors occur at the combinatorial level rather than at the individual label level. In practical engineering applications, if the goal is to identify certain key transmitters or determine the presence of a single transmitter, the evaluation metrics can focus on $P_c^{\text{Hamming}}$, allowing for more robust and easily implementable identification performance.

\section{CONCLUSION}
\label{sec:CONCLUSION}

This paper proposed a SMEI scheme via multi-label learning to effectively identify multiple transmitting devices in overlapping signals. By designing a multi-emitter fingerprint extractor and a classification decision maker, the scheme significantly reduces the complexity while maintaining high accuracy. Experiment results show that the proposed SMEI scheme exhibits excellent robustness and scalability under various SNR and overlap conditions, making it suitable for multi-device concurrent communication scenarios.

\bibliographystyle{IEEEbib}
\bibliography{strings,refs}

\end{document}